\documentclass[twocolumn]{aastex61}
\received{October 19, 2019}
\revised{November 28, 2019}
\accepted{\today}
\submitjournal{ApJ}

\shorttitle{
AGN obscuring fraction implied by SN and radiative feedbacks
}
\shortauthors{Kawakatu, Wada and Ichikawa}

\begin{document}

\title{
Obscuring fraction of active galactic nuclei implied by supernova 
and radiative feedbacks
}

\correspondingauthor{Nozomu Kawakatu}
\email{kawakatsu@kure-nct.ac.jp}

\author{Nozomu Kawakatu}
\affiliation{Faculty of Natural Sciences, 
National Institute of Technology, Kure College, 2-2-11 
Agaminami, Kure, Hiroshima 737-8506, Japan}

\author{Keiichi Wada}
\affiliation{Graduate School of Science and Engineering, Kagoshima University, 
Kagoshima 890-8580, Japan
}
\affiliation
{Ehime University, Research Center for Space and 
Cosmic Evolution, Matsuyama 790-8577, Japan
} 
\affiliation{Hokkaido University, Faculty of Science, 
Sapporo 060-0810, Japan
}

\author{Kohei Ichikawa}
\affiliation{Frontier Research Institute for Interdisciplinary Sciences, Tohoku University, 
Sendai 980-8578, Japan
}
\affiliation{Astronomical Institute, Tohoku University, Sendai 980-8578, Japan
}



\begin{abstract} 
We study the obscuring structure of circumnuclear disks (CNDs) by considering supernova (SN) feedbacks 
from nuclear starburst and the effect of anisotropic radiative pressure from AGNs. 
We suppose that the mass accretion onto a central supermassive black hole (SMBH) is triggered by SN-driven turbulence within CNDs, 
and we explore how the structures of CNDs depend on the BH mass ($M_{\rm BH}$) and AGN luminosity ($L_{\rm AGN}$). 
We find that the obscuring fraction ($f_{\rm obs}$)  peaks at $\sim10\%$ of the Eddington luminosity ($L_{\rm Edd}$), and  its maximal value 
is $f_\mathrm{obs} \sim 0.6$ for less massive SMBHs (e.g., $M_{\rm BH} < 10^{8}M_{\odot}$). This is because the scale height of CNDs is 
determined by the SN-driven accretion for a smaller $L_{\rm AGN}$, while the dusty molecular gas in CNDs is blown away by the radiation 
pressure from AGNs beyond the critical luminosity. On the other hand, 
for massive SMBHs (e.g., $M_{\rm BH} > 10^{8}M_{\odot}$), $f_{\rm obs}$ is always smaller than $0.2$, and it is almost 
independent of $L_{\rm AGN}$ because the scale height of CNDs is mainly controlled by the maximal star-formation efficiency ($C_{\rm *, max}$) 
in CNDs. 
By comparison with the obscuring fractions suggested from the mid-infrared observations of nearby AGNs, the SN plus radiative feedback model 
with  $C_{\rm *, max} = 10^{-7}\, {\rm yr}^{-1}$ well reproduces the observations for  $M_\mathrm{BH} = 10^8 M_\odot$. 
We also find that the intense starburst or the existence of dust-free absorbers inside CNDs are necessary, 
to explain X-ray observations.

\end{abstract}




\section{Introduction} \label{sec:intro}

In the unified model of active galactic nuclei (AGNs) (e.g., Antonucci 1993; Urry \& Padovani 1995),  
supermassive black holes (SMBHs) are obscured by optically and geometrically thick material, 
i.e., the dusty torus.  Recently, ALMA(Atacama Large Millimeter/submillimeter Array) resolved molecular tori of 
a scale of tens of parsecs in nearby Seyfert galaxies 
(Garcia-Burillo et al. 2016; Gallimore et al. 2016; Imanishi et al. 2016, 2018; Izumi et al. 2018; Combes et al. 2019; 
Impellizzeri et al 2019; Garcia-Burillo et al. 2019).   
Because of the anisotropic structures of the tori, the radiation from the nucleus is obscured for particular solid angles.
This obscuring fraction of AGNs (hereafter denoted as $f_{\rm obs}$) should be related to the morphology, size, and 
clumpiness (or internal structure) of their circumnuclear disks, CNDs (e.g., Wada 2015).
In other words, $f_{\rm obs}$ is a key quantity to understand the physical properties of CNDs in AGNs. 
It is also important to understand its cosmological evolution as a function of the BH mass and AGN luminosity 
(or the mass accretion rate) not only to interpret observations of high-$z$ quasars (e.g., Glikman et al. 2011; Ikeda et al. 2011; 
Ikeda et al. 2012; Masters et al. 2012; McGreer et al. 2013, 2018; Yang et al. 2016; Matsuoka et al. 2018) but also to improve 
theoretical models for the evolution of AGNs (e.g., Fanidakis et al. 2012; Enoki et al. 2014; Lapi et al. 2014; Shirakata et al. 2019).

Statistical studies suggested that $f_{\rm obs}$ depends on AGN properties, such as the AGN luminosity ($L_{\rm Edd}$), 
and the results are not fully consistent among studies using samples with different wavelengths. 
The fraction of type-2 AGNs inferred from the infrared observations, or the infrared-to-bolometric luminosity ratio, depends on the AGN luminosity 
 (e.g., Maiolino et al. 2007; Treister et al. 2008; Alonso-Herrero et al. 2011, Lusso et al. 2013; Toba et al. 2013; Toba et al. 2014; Ichikawa et al. 2017, 2019). 
The obscuring fraction $f_{\rm obs}$ also affects on the classification of Seyfert galaxies as a type-1 or type 2  
(e.g., Alonso-Herrero A. et al., 2011; Ramos Almeida C. et al., 2011; Ichikawa et al. 2015; Audibert et al.,  2017). 
They found that type 2 Seyfert galaxies  require higher extinction values and larger covering factor than that 
for type 1 Seyfert galaxies. 
X-ray observations also suggest that the fraction of obscured Compton-thin AGNs clearly decrease with the AGN luminosity 
(e.g., Ueda et al. 2003; La Franca et al. 2005; Hasinger 2008; Merloni et al. 2014; Ricci et al. 2014; Ueda et al. 2014; Akylas et al. 2016), 
which may be interpreted as a simple receding torus model (e.g., Lawrence 1991; Simpson 2005). 
Recent X-ray studies suggested that $f_{\rm obs}$ is also smaller for less luminous AGNs, and the obscured properties could also be different 
in high-$z$ AGNs (e.g., Burlon et al. 2011: Buchner et al. 2015; Kawamuro et al. 2016; Buchner \& Bauer 2017). 
A more physics-motivated study based on radiation-hydrodynamic models succeed in explaining these observations to some extent (Wada 2015). 
On the other hand, Ricci et al. (2017) suggested that $f_{\rm obs}$ depends mostly on the Eddington luminosity ratio ($L_{\rm AGN}/L_{\rm Edd}$), 
rather than the AGN luminosity, based on a survey using the all-sky hard X-ray $\it{Swift}$ Burst Alert Telescope, where 
$L_{\rm Edd}=4\pi cGM_{\rm BH}m_{\rm p}/\sigma_{\rm T}$. 
These observations indicate more complicated behavior in terms of the BH mass, AGN luminosity, and Eddington luminosity ratio, 
but the physical mechanisms behind them remain unclear.

A key physical phenomenon to understand the properties of the obscuration in the circumnuclear region is the effect of star formation. 
In fact, prominent star formation has been observed in the central sub-kpc regions of nearby AGNs
 (e.g., Imanishi \& Wada 2004; Davies et al. 2007; Imanishi et al. 2011; Diamond-Stanic \& Rieke 2012; Hicks et al. 2013; Davies et al. 2014; 
Alonso-Herrero et al. 2014; Esquej et al. 2014; Mallmann et al. 2018). 
Recently, Izumi, Kawakatu, \& Kohno (2016) found a positive correlation between the mass of dense molecular gas in CNDs of 
the scale of $\sim 100$ pc and the mass accretion rate onto an SMBH. Thus, these findings suggest that  nuclear star formation 
may be related to both the formation of the AGN obscuring structure and AGN activity. 
On the stellar population in the vicinity of AGNs, recent near-infrared IFU (integral field unit) observations have revealed 
the presence of young to intermediate age stars 
(e.g. Riffel et al. 2007; Riffel et al. 2009; Riffel et al. 2010; Riffel et al. 2011; Storchi-Bergmann et al. 2012; Ruschel-Dutra et al. 2017; 
Hennig et al. 2018; Diniz et al. 2019). 
It indicates the possibility of large amounts of type-II supernova (SN) in the central region of AGNs. 
Along these lines, we proposed a simple model of a nuclear starburst disk supported by the turbulent pressure from type II SN explosions
(Kawakatu \& Wada 2008: hereafter KW08; Kawakatu \& Wada 2009), because our main aim is to clarify how the obscuring structure depends on 
physical quantities of AGNs and CNDs by changing a wide range of physical parameters. 
There also exist numerous theoretical and numerical models of AGN obscuring  structures; 
e.g., (1) radiation pressure from AGNs (e.g., Krolik 2007; Namekata \& Umemura  2014, 2016; Williamson et al. 2019), 
(2) radiation pressure from nuclear starburst (e.g., Ohsuga \& Umemura 1999; Thompson et al. 2005), 
(3) high-velocity dispersion clouds/clumps (e.g., Krolik \& Begelman 1988; Vollmer  et al. 2008), 
(4) turbulent pressure from  type-II  SN explosions (e.g., Wada \& Norman 2002; Wada et al. 2009), 
(5) disk winds (e.g., Elitzur \& Shlosman 2006; Nomura et al. 2016, 2017) , 
(6) radiation-induced warping disks (e.g., Pringle 1997), and 
(7) outflows driven by  AGN radiation pressure (Wada 2015; Wada et al. 2016; Dorodnitsyn et al. 2016; Chan \& Krolik 2016, 2017). 
Herein, we study the effect of radiative feedback as a key physical phenomenon to determine
 the obscuring structure of AGNs. 

In this paper, we extend KW08 to investigate the obscuring fraction of AGNs $f_{\rm obs}$ due to 
the absorption of dusty gas  in CNDs of the scale of $1$--$10$~pc  by taking account of the anisotropic radiation pressure 
from AGNs. In particular, we explore how $f_{\rm obs}$ depends on the BH mass, AGN luminosity, and physical properties of CNDs.
We then compare the theoretical models with observationally suggested obscuring fractions.

The remainder of this paper is organized as follows. In $\S 2$, we summarize both the SN-driven turbulence model and the radiative feedback model. 
We show the predicted AGN obscuring fraction $f_{\rm obs}$ and how $f_{\rm obs}$ depends on the BH mass, AGN luminosity, and other 
physical quantities of CNDs in $\S 3$. We compare these theoretical results with IR and X-ray observations in $\S 4$. 
Finally, a summary is presented in $\S 5$.

\section{Models}
Based on KW08, in which a CND supported by the turbulent pressure from SN explosions was studied, 
we evaluate the obscuring fraction $f_{\rm obs}$ (or covering angle $\frac{\pi}{2}-\theta_{\rm CND}$). 
The obscuring fraction $f_{\rm obs}$ is defined as 
\begin{eqnarray}
f_{\rm obs}=\frac{1}{2\pi}{\int_{0}^{2\pi}\int_{\theta_{\rm CND}}^{\pi/2}\sin{\theta}d\theta d\phi}=\cos{\theta_{\rm CND}}.
\end{eqnarray}
Here, $\theta_{\rm CND}$ is the maximal thickness of CNDs, i.e., $\tan\left({\frac{\pi}{2} -\theta_{\rm CND}}\right.)
=h(r_{\rm out})/r_{\rm out}$, 
as schematically shown in Fig. 1, where $h$ and $r_{\rm out}$ are the scale height and outer radius of CNDs, respectively.  
We here assume that the scale height is determined by  SN-driven turbulence (section 2.1), 
following the analytical study by  KW08. 
In section 2.2, we additionally  consider the effect of anisotropic radiation pressure from AGNs.

\subsection{SN-driven turbulent disk}
We assume that the vertical structure of CNDs is in hydrostatic equilibrium (see details in Wada \& Norman 2002). 
The turbulent pressure associated with SN explosions is 
balanced with gravitation in the vertical direction by 
\begin{equation}
\rho_{\rm g}v_{\rm t}^{2} = \rho_{\rm g}gh, 
\end{equation}
where $\rho_{\rm g}$, $v_{\rm t}$, and $h$ are the gas density, 
turbulent velocity, and scale height of the disk, respectively. 
In the region where the gravity of SMBH dominates, the $z$ direction of gravity, $g$, is obtained as 
$g\equiv GM_{\rm BH}h/r^{3}$, where $r$ is the radial distance from a central BH.
We assume that the turbulence is driven by the energy input from SN explosions.  
The energy loss $E_{\rm out}$ due to the turbulent dissipation is given by
\begin{equation}
E_{\rm out}=\frac{\rho_{\rm g}v_{\rm t}^{2}}{t_{\rm dis}}
=\frac{\rho_{\rm g}v_{\rm t}^{3}}{h},
\end{equation}
where the dissipation timescale of the turbulence 
$t_{\rm dis}=h/v_{\rm t}$. 

The energy input from SN explosions, $E_{\rm in}$, can be expressed as
\begin{equation}
E_{\rm in}=\epsilon_{\rm SN}f_{\rm SN}E_{\rm SN}S_{*}, 
\end{equation}
where $E_{\rm SN}$ is the total energy ($10^{51}\,{\rm erg}$) injected by an SN; 
$S_{*}$ is the star-formation rate per unit volume and time; and 
$\epsilon_{\rm SN}$ and $f_{\rm SN}$ are the efficiency with which the SN energy 
is transferred to the gas in the CND  and the number density of supernovae (SNe) per solar 
mass of the star formation, respectively. 
In this paper,  we assume that $\eta \equiv \epsilon_{\rm SN} f_{\rm SN} =10^{-3}M^{-1}_{\odot}$ with $\epsilon_{\rm SN}=0.1$
 (e.g., Thornton et al. 1998; Wada \& Norman 2002; Wada et al. 2009), and $f_{\rm SN}=10^{-2}M^{-1}_{\odot}$,  
which is expected for the Salpeter initial mass function (IMF) with the low-mass cutoff $m_{\rm l}=0.1M_{\odot}$
(e.g.,  Thompson et al. 2005)\footnote{Note that the detection of SNe at the center of galaxies is quite hard 
because the huge column densities around AGNs cause enormous extinction, and a high angular resolution 
is needed to discern individual SNe. Thus far, the radial distribution of SNe in the central galactic region has been 
analyzed for only a few starburst galaxies (e.g., Herrero-Illana et al. 2012).}. 

Under the energy balance $E_{\rm in}=E_{\rm out}$, we obtain 
\begin{equation}
\frac{\rho_{\rm g}v_{\rm t}^{3}}{h}
=\eta E_{\rm SN}C_{*}\rho_{\rm g}. 
\end{equation}
Here, we assume a star-formation recipe $S_{*}=C_{*}\rho_{\rm g}$, where $C_{*}$ is the star-formation efficiency. 
Using eqs. (2), (3), and (5), the turbulent velocity $v_{\rm t}$ and scale height $h$ are expressed as 
\begin{eqnarray}
v_{\rm t} &=& \left(\frac{GM_{\rm BH}}{r^{3}}\right)^{1/2}h, \\
h &=& \left(\frac{GM_{\rm BH}}{r^{3}}\right)^{-3/4}(\eta E_{\rm SN}C_{*})^{1/2}, \nonumber \\
{} &=& 14\, {\rm pc} \left(\frac{C_{*}}{10^{-8}\,{\rm yr}^{-1}}\right)^{1/2}
\left(\frac{M_{\rm BH}}{10^{7}M_{\odot}}\right)^{-3/4}
\left(\frac{r}{30\,{\rm pc}}\right)^{9/4}.
\end{eqnarray}
From eq. (7), the SN-driven turbulence model predicts a concave structure for CNDs, 
i.e., $h\propto r^{9/4}$ (see Fig. 1). 
The turbulent velocity is re-written as follows: 
\begin{eqnarray}
v_{\rm t} &=& \left(\frac{GM_{\rm BH}}{r^{3}}\right)^{-1/4}(\eta E_{\rm SN}C_{*})^{1/2}, \nonumber \\
{} & = & 18\, {\rm km}\, {\rm s}^{-1} \left(\frac{C_{*}}{10^{-8}\,{\rm yr}^{-1}}\right)^{1/2}
\left(\frac{M_{\rm BH}}{10^{7}M_{\odot}}\right)^{-1/4}
\left(\frac{r}{30\,{\rm pc}}\right)^{3/4}.
\end{eqnarray}
Thus, the turbulent velocity increases with the star-formation efficiency and decreases with the BH mass 
for a given $r$. This trend is consistent with observations for nearby Seyfert galaxies (e.g., Hicks et al. 2009). 

The star-formation efficiency $C_{*}$ is related to the star-formation mode (i.e., normal/starburst), 
formation redshift  (low-$z$/high-$z$), and formation sites (bars and spiral arms), 
which are supported by numerous theoretical and observational studies (e.g., Komugi et al. 2005; Bigiel et al. 2008; 
Wada \& Norman 2007; Dobbs \& Pringle 2009; Krumholz et al. 2009,  Daddi et al. 2010; Momose et al. 2010). 
Thus, we here consider a wide range of the star-formation efficiency
$C_{*}$, i.e., $10^{-10}\,{\rm yr}^{-1} \leq C_{*} \leq  10^{-6}\,{\rm yr}^{-1}$ (see also Fig. 5 in Kawakatu \& Wada 2009). 
If the maximum star-formation efficiency $C_{*,{\rm max}}$ is assumed to be $10^{-7}\,{\rm yr}^{-1}$ as the fiducial case, 
the upper limit of the thickness of CNDs is obtained as 
\begin{eqnarray}
\frac{h(r_{\rm out})}{r_{\rm out}}=\tan{\left(\frac{\pi}{2}-\theta_{\rm CND} \right)} 
\simeq & 1.5 \left(\frac{C_{\rm *, max}}{10^{-7}\,{\rm yr}^{-1}}\right)^{1/2} \nonumber \\
\left(\frac{M_{\rm BH}}{10^{7}M_{\odot}}\right)^{-3/4}
\left(\frac{r_{\rm out}}{30\,{\rm pc}}\right)^{5/4}. 
\end{eqnarray}
This indicates that the CND thickness increases with the star-formation efficiency 
and decreases with the BH mass.
In section 3.1, we will  investigate the relation between the obscuring fraction (the thickness of CNDs) 
and the AGN luminosity.

For the inner radius of CNDs ($r_{\rm in}$), since the SN-turbulence model works as far as  the star formation occurs in CNDs, 
$r_{\rm in}$ is not necessarily determined by the dust sublimation radius, $r_{\rm sub}$. Considering anisotropic radiation from AGNs, 
$r_{\rm sub}$  is not sharp boundary and closer  to  the  central  black  hole (e.g., Kawaguchi \& Mori 2010).
Izumi, Kawakatu \& Kohno (2016) estimated the inner radius as $r_{\rm in}=\max[{r_{\rm X}, r_{\rm Q}}]$. 
Here, $r_{\rm X}$ is the radius at which the ratio of the X-ray energy-deposition rate ($H_{\rm X}$) and gas number density ($n_{e}$) takes $\log(H_{\rm X}/n_{e}) =-27.5$.
Note that, in the region with $\log(H_{\rm X}/n_{\rm e}) =-27.5$, 
the gas temperature is approximately 100 K (Maloney et al. 1996). 
On the other hand, $r_{\rm Q}$ is the radius which is determined by Toomre's stability criterion (Toomre \& Toomre 1972), i.e., when the surface density 
of the gas in the CND, $\Sigma_{\rm g}$, is higher  than the critical surface density, $\Sigma_{\rm crit}$,  the CND is gravitationally unstable. Following KW08, 
the critical surface density at $r_{\rm in}$ is given by 
\begin{eqnarray}
\Sigma_{\rm g, crit}(r_{\rm in}) \simeq 3.4\,{\rm g\, cm^{-2}}
\left(\frac{c_{\rm s}}{1\,{\rm km\,s^{-1}}}\right) \nonumber \\
\left(\frac{r_{\rm in}}{1\,{\rm pc}} \right)^{-3/2} 
\left(\frac{M_{\rm BH}}{10^{7}M_{\odot}}\right)^{1/2}. 
\end{eqnarray}
Because $c_{\rm s}=(5kT_{\rm g}/3m_{\rm p})^{1/2}$, where $k$ and $m_{\rm p}$ are the Boltzmann constant and proton mass, respectively, 
the corresponding gas temperature is $T_{\rm g}=100\, {\rm K}$.  
Here, we assume that an isothermal cold gas dominates the mass ($T_{\rm g}=50-100\,{\rm K}$) in CNDs because the molecular and dust cooling is effective 
(e.g., Wada \& Tomisaka 2005; Wada et al. 2009). 
On apply this concept to nearby Seyfert galaxies, 
Izumi et al. (2016) found that the range of $r_{\rm in}$ is $0.1 - 2$ pc (see Table 4 in Izumi et al. 2016), which is consistent with the results derived by 
the comparison of the infrared nuclear spectral energy distributions (SEDs) with the {\it CLUMPY} torus model 
(e.g., Alonso-Herrero A. et al., 2011; Ramos Almeida C. et al., 2011; Ichikawa et al. 2015; Audibert et al.,  2017). 
Thus, we here assume the inner radius of CNDs as $r_{\rm in}=1\,{\rm pc}$ for the fiducial case. 
We will discuss the dependence of $r_{\rm in}$ in $\S 3.4$.  
The outer radius $r_{\rm out}$ is defined as the outer boundary inside which the potential of the BH dominates 
that of CNDs. Thus,  $r_{\rm out}$ is given by 
\begin{eqnarray}
r_{\rm out} &=& \left(\frac{M_{\rm BH}}{\pi \Sigma_{g}}\right)^{1/2} \nonumber \\
{} & = & 30\,{\rm pc} \left(\frac{M_{\rm BH}}{10^{7}M_{\odot}}\right)^{1/2}
\left(\frac{\Sigma_{\rm g}}{1\,{\rm g}\,{\rm cm}^{-2}}\right)^{-1/2}, 
\end{eqnarray}
where $\Sigma_{\rm g}$ is the surface density of CNDs.  
This radius is comparable to the radius of the dusty torus of NGC~1068 (Garcia-Burillo et al. 2016 
Imanishi et al. 2016, 2018; Garcia-Burillo et al. 2019) and the Circinus galaxy (Izumi et al. 2018). 

\subsection{Effect of AGN radiative feedbacks}
In order to examine how the radiation pressure from AGNs (i.e., accretion disk) affects the structure of CNDs 
predicted by the SN-driven turbulent disk ($\S 2.1$), 
we consider anisotropic radiation from an AGN emitted by an accretion disk around a SMBH, 
following previous work (e.g., Netzer 1987; Kawaguchi \& Mori 2010, 2011; Liu \& Zhang 2011; 
Namekata \& Umemura 2016). In this section, we evaluate the obscuring fraction, $f_{\rm obs}$, predicted 
by the model that takes into account  not only the SN feedback but also the radiative feedback from the AGN 
(hereafter, we call it the hybrid model).

The radiation force from AGN, $F_{\rm rad}$, is obtained as 
\begin{equation}
F_{\rm rad}(\theta)=\frac{\chi_{\rm d}}{c}\frac{6}{7}\frac{L_{\rm AGN}}{4\pi r^{2}}\frac{1-e^{-\bar{\tau}}}{\bar{\tau}}
\cos{\theta}(1+2\cos{\theta}), 
\end{equation}
where $\chi_{\rm d}$, $\bar{\tau}$, and $\theta$ are the mass extinction of dusty gas, the optical depth of clumpy clouds 
(i.e., the average optical depth of line of sight),  and the angle between the line of sight and the normal of the accretion disk (see Fig. 1). 
Here, we assume that the CND is alighted with the accretion disk. 
Note that the orientations of accretion disk may be possible independent of the CND (e.g., Kawaguchi \& Mori 2010). 
If this is the case, $\theta$-dependence of $F_{\rm rad}(\theta)$ would be relatively weak but this effect does not change our 
main results (see Wada 2015).

\section{Results}
First, we derive the obscuring fraction $f_{\rm obs}$ predicted by the SN-driven turbulence model in section 3.1.  
In section 3.2, we also examine the effect of anisotropic radiation pressure from AGNs on $f_{\rm obs}$ based on the model described in \S 3.1. 
In section 3.3, we explore how the obscuring fraction depends on the BH mass and AGN luminosity. Finally, in section 3.4, we discuss the 
dependence on the physical parameters of CNDs (the inner radius $r_{\rm in}$, the surface density $\Sigma_{\rm g}$ 
and the average optical depth of line of sight $\bar{\tau}$). 

\subsection{Obscuring fraction in an SN-driven turbulent disk}
We assume a kinetic viscosity, expressed as follows, as a source of angular-momentum transfer in the gas: 
$\nu_{\rm t}=\alpha_{\rm SN} v_{\rm t}h$, where $\alpha_{\rm SN}$ ($\leq 1$) is a parameter. 
Hereafter, we assume that $\alpha_{\rm SN}=1$, which is supported by numerical simulations (e.g., Wada \& Norman 2002). 
The mass accretion rate in a viscous accretion disk is then given by 
\begin{eqnarray}
\dot{M}_{\rm acc}(r) &=& 2\pi \, \nu_{\rm t}\, \Sigma_{\rm g}(r)\, \left|\frac{d\,{\rm ln}\,\Omega_{\rm K}(r)}{d\,{\rm ln}\, r}\right|, 
\end{eqnarray}
where $\Omega_{\rm K}(r)$ is the angular velocity in the Kepler motion, 
i.e.,  $\Omega_{\rm K} (r)=\left(GM_{\rm BH}/r^{3}\right)^{1/2}$. 

Assuming $\Sigma_{\rm g}=\Sigma_{\rm g, crit}$, i.e., marginally unstable, the mass accretion rate 
at the inner radius $r_{\rm in}$ can be expressed  as   
\begin{eqnarray}
\scalebox{0.9}{$\displaystyle
\dot{M}_{\rm acc}(r_{\rm in}) $}&=& 
\scalebox{0.9}{$\displaystyle
3\pi \, \eta \,  E_{\rm SN}\, C_{*}\, \Sigma_{\rm g}(r_{\rm in})
\left(\frac{{r_{\rm in}}^{3}}{GM_{\rm BH}}\right) $}, \\
{}&=&\scalebox{0.9}{$\displaystyle
0.005 \left(\frac{r_{\rm in}}{1\,{\rm pc}}\right)^{3}
\left(\frac{C_{\rm *, max}}{10^{-7}\,{\rm yr}^{-1}}\right) $}\nonumber \\
{} &\times& \scalebox{0.9}{$\displaystyle
\left(\frac{\Sigma_{\rm g,crit}}{1\,{\rm g\, cm^{-2}}}\right) 
\left(\frac{M_{\rm BH}}{10^{7}M_{\odot}}\right)^{-1} M_{\odot}\,{\rm yr}^{-1}
$}. 
\end{eqnarray}

Assuming $r_{\rm out}=10\,{\rm pc}$, we obtain $\dot{M}_{\rm acc}(r_{\rm in})/\dot{M}_{*} \sim 0.1\,(M_{\rm BH}/
10^{7}M_{\odot})^{-1}\,(r_{\rm out}/10\,{\rm pc})^{-2}$, where the star formation rate is $\dot{M}_{*}=C_{\rm *, max}
\,\Sigma_{\rm g, crit}\,r_{\rm out}^2$. This is consistent with the observations that indicate a close connection between 
AGN and the nuclear starburst (e.g., Imanishi \& Wada 2004; Diamond-Stanic \& Rieke 2012; Alonso-Herrero et al. 2014; Esquej et al. 2014). 
Our model also explains the correlation between the dense gas mass of CNDs and the AGN luminosity for nearby Seyfert galaxies 
(Fig.3 in Izumi, Kawakatu \& Kohno 2016) and nearby radio galaxies NGC 1275 (Nagai et al. 2019).  

Although the growth rate of SMBHs, i.e., $\dot{M}_{\rm BH}$, is not necessarily equal to the mass accretion rate at the inner boundary, 
$\dot{M}(r_{\rm in})$, we here assume the maximal mass accretion rate, 
i.e., $\dot{M}_{\rm acc}(r_{\rm in})= \dot{M}_{\rm BH}$, we can estimate the AGN bolometric luminosity 
because $L_{\rm AGN}$ is given as a function of $\dot{M}_{\rm BH}/\dot{M}_{\rm Edd}$ (Watarai et al. 2000):
\begin{equation}
L_{\rm AGN}=\left \{
 \begin{array}{l}
 2\left(1+\ln{\frac{\dot{M}_{\rm BH}/\dot{M}_{\rm Edd}}{20}}\right)
L_{\rm Edd}\,\,\, ;\dot{M}_{\rm BH}/\dot{M}_{\rm Edd} \geq 20, \\ \\
 \left(\frac{\dot{M}_{\rm BH}/\dot{M}_{\rm Edd}}{10}\right)
L_{\rm Edd} \,\,\,\,\, ;\dot{M}_{\rm BH}/\dot{M}_{\rm Edd} < 20, 
 \end{array}\right .
\end{equation}
where $\dot{M}_{\rm Edd}=L_{\rm Edd}/c^{2}$ is the Eddington mass accretion rate. 
$L_{\rm Edd}=4\pi cGM_{\rm BH}m_{\rm p}/\sigma_{\rm T}$, where 
$m_{\rm p}$ and $\sigma_{\rm T}$ are the proton mass and Thomson cross section, respectively.

We rewrite eq. (9)  using eqs. (15) and (16) as
\begin{eqnarray}
\frac{h(r_{\rm out})}{r_{\rm out}}=\tan{\left(\frac{\pi}{2}-\theta_{\rm CND} \right)} 
&=& 1.0 \left(\frac{\dot{M}_{\rm BH}}{\dot{M}_{\rm Edd}} \right)^{1/2}
\left(\frac{\Sigma_{\rm g}(r_{\rm in})}{\Sigma_{\rm g, crit}(r_{\rm in})}\right)^{-1/2} \nonumber \\
\left(\frac{r_{\rm in}}{1\,{\rm pc}}\right)^{-3/4}
\left(\frac{r_{\rm out}}{30\,{\rm pc}}\right)^{5/4}.
\end{eqnarray}
Since $f_{\rm obs}$ depends on $\theta_{\rm CND}$ (see eq. (1)), by combing with eqs. (16) and (17),  
$f_{\rm obs}$ can be obtained as  a function of the Eddington ratio $L_{\rm AGN}/L_{\rm Edd}$. 
Here, we note that the scale height ($h/r$) at the galactic scale ($r > r_{\rm out}$) is smaller than 
that at the CND scale because $h/r \propto v_{\rm t}/v_{\phi}$, where $v_{\phi}$ is the circular velocity.   
(see Wada \& Norman 2002). 

In Fig. 2, the blue dashed line shows the covering angle ($\frac{\pi}{2} -\theta_{\rm CND}$) and $f_{\rm obs}$ 
as functions of $L_{\rm AGN}$ predicted by the SN-driven turbulence model for the typical BH mass of Seyfert galaxies 
with $M_{\rm BH}=10^{7}M_{\odot}$ (e.g., Wu, \& Han 2001). 
Given a  BH mass,  the horizontal dashed line is plotted as a covering angle (or $f_{\rm obs}$) for the maximal star-formation efficiency 
$C_{\rm *, max}=10^{-7}\,{\rm yr}^{-1}$ (see eq. (9)) .
We find that the obscuring fraction $f_{\rm obs}$ monotonically increases with $L_{\rm AGN}$ because the covering angle increases 
as $\dot{M}_{\rm BH}$ increases (see eq.(17)). 
On the other hand, the maximal value of $f_{\rm obs}$ follows the horizontal dashed line determined by the maximal star-formation efficiency,  
 $f_{\rm obs}\, (C_{\rm *, max})$ (see eq. (9)). The maximum value of $f_{\rm obs}$ is $\sim 0.8$ for $L_{\rm AGN}\ge 0.1L_{\rm Edd}$.

\subsection{Obscuring fraction with radiative feedback}
When the gravitational force, $F_{\rm grav}=GM_{\rm BH}/r^{2}$, is balanced by 
the anisotropic radiation force, $F_{\rm rad}$ (see eq. (12)), we can obtain the critical angle $\theta_{\rm crit}$ by using 
the total luminosity of the AGN, $L_{\rm AGN}$, which is defined in eq. (15). 
A part of the dusty torus corresponding to $\theta_{\rm CND} < \theta_{\rm crit}$ is blown away by the radiation pressure. 
The critical angle $\theta_{\rm crit}$ is obtained as 
\begin{equation}
\cos{\theta_{\rm crit}}\left(1+2\cos{\theta_{\rm crit}}\right)=\frac{7}{6A}\left(\frac{L_{\rm AGN}}{L_{\rm Edd}}\right)^{-1}, 
\end{equation}
where the boost factor $A=(\chi_{\rm d}/\chi_{\rm T})(1-e^{-\bar{\tau}}/\bar{\tau})$. $\chi_{\rm T}=\sigma_{\rm T}/m_{\rm p}$, 
where $\sigma_{\rm T}$ and $m_{\rm p}$ are the Thomson cross-sectional area and proton mass, respectively.   
We assume $\chi_{\rm d}=100\,{\rm cm}^{2}\,{\rm g}^{-1}$ and the optical depth of line of sight with $\bar{\tau}=10$ 
as a fiducial case, but  we will examine the dependences of $\bar{\tau}$ on $f_{\rm obs}$ in $\S 3. 4$.

In Fig.3,  the red dashed line shows the effect of the radiation pressure on $f_{\rm obs}$ (see eq. (18)). 
The thick black line represents the obscuring fraction $f_{\rm obs}$ predicted by the hybrid model, which considers 
both the SN feedback and the radiative feedback from the AGN. 
As a result, the obscuring fraction $f_{\rm obs}$ peaks at approximately $10\% $ of the Eddington luminosity, $L_{\rm AGN, p}\sim 0.1L_{\rm Edd}$, 
and  its maximum value is $\sim 0.6$, 
which is comparable to the type-2 fraction of nearby Seyfert galaxies (e.g., Roseboom et al. 2013; Lusso et al. 2013; Shao et al. 2013).
As $L_{\rm AGN} < L_{\rm AGN,p}$, the obscuring fraction increases with $L_{\rm AGN}$ because the SN feedback  is more effective 
than the radiative feedback. 
On the other hand,  when $L_{\rm AGN} > L_{\rm AGN,p}$, $f_{\rm obs}$ decreases with increasing $L_{\rm AGN}$ owing to the radiation pressure from AGNs.

\subsection{Dependences on BH mass and AGN luminosity} 
Here, we investigate the dependence of the obscuring fraction $f_{\rm obs}$ on $M_{\rm BH}$  and  $L_{\rm AGN}$ 
by assuming $r_{\rm in}=1\,{\rm pc}$, $\Sigma_{\rm g}=\Sigma_{\rm g, crit}$, and $\bar{\tau}=10$. 
Based on the argument in previous sections, the obscuring fraction ($f_{\rm obs}$) is plotted as a function of $M_{\rm BH}$ and $L_{\rm AGN}$ in Fig. 4 
and the Eddington ratios $\lambda_{\rm Edd}=L_{\rm AGN}/L_{\rm Edd}$ for various BH masses in Fig. 5.
These show that the obscuring fraction strongly depend on the Eddington ratio ($\lambda_{\rm Edd}$) for smaller BHs 
($M_{\rm BH}< 10^{8} M_{\odot}$);  it is largest for $\lambda_{\rm Edd}\sim 0.1$.
For more massive BHs ($M_{\rm BH} > 10^{8} M_{\odot}$), $f_{\rm obs}$ weakly depends on the Eddington ratio. 
Thus, it seems that the behavior of $f_{\rm obs}$ changes around the typical BH mass, $M_{\rm BH, t}\simeq 10^{8}M_{\odot}$. 
The typical BH mass is determined by the equations of the maximal obscuring fractions of the hybrid model, $f_{\rm obs}$ (eqs. (17) and (18))
and $f_{\rm obs} \, (C_{*, {\rm max}})$ (eq. (9)), as follows:  
\begin{equation}
M_{\rm BH, t}\simeq 8\times 10^{7}M_{\odot}\left(\frac{C_{\rm *, max}}{10^{-7}\,{\rm yr}^{-1}} \right) ^{2/3}
\left(\frac{L_{\rm AGN, p}/L_{\rm Edd}}{0.03} \right)^{-2/3}, 
\end{equation}
where $L_{\rm AGN, p}/L_{\rm Edd}\simeq 0.03$ for $\bar{\tau}=10$, as shown in Fig. 3. 
Note that $M_{\rm BH, t}$ becomes smaller because $L_{\rm AGN, p}/L_{\rm Edd}$ increases with $\bar{\tau}$ (see Fig. 8). 
Figure 4 and Figure 5 also show that both AGNs with higher Eddington ratios ($L_{\rm AGN}/L_{\rm Edd} > 1$) and 
those with lower Eddington rations ($L_{\rm AGN}/L_{\rm Edd} < 10^{-2}$) are surrounded by 
geometrically thin CNDs (small $f_{\rm obs}$) owing to the strong outflow driven by the radiation pressure from AGNs 
and lower star-formation efficiency $C_{*}$, respectively.

For less massive BHs ($M_{\rm BH} < M_{\rm BH, t}$), the relation between $f_{\rm obs}$ and $L_{\rm AGN}$ is similar to 
that for $M_{\rm BH}=10^{7}M_{\odot}$. The only difference is that the maximal $f_{\rm obs}$ is slightly smaller
because the outer radius decreases as the  BH mass decreases [i.e., $r_{\rm out}\propto M_{\rm BH}^{1/2}$ (see eq. (11))]. 
Thus, $\tan{\left(\frac{\pi}{2}-\theta_{\rm CND} \right)} $ (or the maximal $f_{\rm obs}$) becomes small (see eq.(17)). 
In fact, we found that the maximal $f_{\rm obs}\simeq 0.4$ for $M_{\rm BH} = 10^{6}M_{\odot}$ 
(cf. maximal $f_{\rm obs}\simeq 0.6$ 
for $M_{\rm BH} = 10^{7}M_{\odot}$ ).

For more massive BHs ($M_{\rm BH} > M_{\rm BH, t}$), the behavior of  $f_{\rm obs}$ with respect to $L_{\rm AGN}$ is 
different from that for the less massive BHs. Figure 4 shows that the obscuring fraction remains at a a low level  (i.e., $f_{\rm obs}  < 0.2 $),  
and $f_{\rm obs}$ weekly depends on $L_{\rm AGN}$. 
In order to reveal the reason, in Fig. 6, we examine how the obscuring fraction depends on $L_{\rm AGN}$ for AGNs with $M_{\rm BH}=10^{8}M_{\odot}$. 
We find that the maximal obscuring fraction is determined by $C_{\rm *,max}$, which is different from the case of $M_{\rm BH} < M_{\rm BH, t}$. 
This is because the upper limit of the obscuring fraction $f_{\rm obs} \, (C_{*, {\rm max}})$ decreases as the BH mass increases (see eq. (9)). 
Thus, the $f_{\rm obs}$ estimated using the hybrid model (blue and red dashed lines) can be greater than $f_{\rm obs} \, (C_{*, {\rm max}})$.  
In particular, as shown in Fig. 4, the dependence on $M_{\rm BH}$ is conspicuous for $M_{\rm BH}=10^{9}M_{\odot}$ because 
$\tan{\left(\frac{\pi}{2}-\theta_{\rm CND} \right)} \propto M_{\rm BH}^{-3/4}r_{\rm out}^{5/4}\propto M_{\rm BH}^{-1/8}$ 
(see eqs. (11) and (17)). Consequently, the dependence of $f_{\rm obs}$ on $L_{\rm AGN}$ is weak. 
Therefore, the present model could explain why the fraction of the type 2 QSO is much smaller than that of Seyfert galaxies.

\subsection{Dependences on physical parameters of CNDs}
There are three free parameters that could affect $f_{\rm obs}$: $r_{\rm in}$, $\bar{\tau}$, and $\Sigma_{\rm g}$.
First, we examine the effect of decreasing the inner radius, i.e., $r_{\rm in} \leq 1$ pc. 
In the inner few hundred parsecs of ultra-luminous infrared galaxies (ULIRGs with the infrared luminosity $L_{\rm IR} > 3.8\times 10^{45}{\rm erg/s}$), 
the average gas number density reaches 
$10^{4}-10^{5}\,{\rm cm}^{-3}$, which is higher than that in normal AGNs (e.g., Thompson et al. 2005; Scoville et al. 2015).  
In this case, star formation may occur at a smaller inner radius.  
Figure 7 compares $f_{\rm obs}$ in two models with $r_{\rm in} = 1\, {\rm pc}$ and $0.3$ pc.
For a smaller $r_{\rm in}$, $f_{\rm obs}$ is larger for any $L_{\rm AGN}$. 
Since $L_{\rm AGN, p}$ decreases as $r_{\rm}$ is smaller, the SN feedback works effectively.   
Consequently, at the peak AGN luminosity ($L_{\rm AGN, p}\sim 0.03L_{\rm Edd}$),  the maximal  obscuring fraction reaches 
$f_{\rm obs}\sim 0.8$, in contrast to 0.6 for $r_{\rm in}=1\,{\rm pc}$. 

Second, we examine how a different $\bar{\tau}$ changes the present results. 
Figure 8 shows that the obscuring fraction $f_{\rm obs}$ is a function of $L_{\rm AGN}$ for $\bar{\tau}=1, \,10$, and $100$. 
As shown in Fig. 8, the peak AGN luminosity ($L_{\rm AGN, p}$) increases and $f_{\rm obs}$ decreases as $\bar{\tau}$ increases because the effect of radiation pressure 
becomes weaker owing to the absorption of denser gas clouds (see eq. (18)). 
Thus, this effect changes the $f_{\rm obs}-\lambda_{\rm Edd}$ relation as seen in Fig. 5, e.g., for $\bar{\tau}=1$,  
the peak Eddington ratio $L_{\rm AGN, p}/L_{\rm Edd}\simeq 0.01$ and the maximal $f_{\rm obs}\simeq 0.3$ 
Interestingly, according to the model fitting of infrared AGN SEDs (e.g., e.g., Alonso-Herrero A. et al., 2011; Ramos Almeida C. et al., 2011; 
Ichikawa et al. 2015; Audibert et al.,  2017), they found that type 2 Seyfert galaxies (Sy2)  require higher extinction values (i.e., higher $\bar{\tau}$) 
and larger covering factor (i.e., higher $f_{\rm obs}$) than that for type 1 Seyfert galaxies (Sy1). This is consistent with our predictions, i.e., $f_{\rm obs}\simeq 0.6$ for $\bar{\tau}=10$ and $f_{\rm obs}\simeq 0.8$ for $\bar{\tau}=10^{2}$ as seen in Fig. 8. However, it is still under debate 
why some Sy2s possess intrinsically higher optical depth, $\bar{\tau}$. This is left in our future work.

Because the optical depth of clouds is $\bar{\tau}=\chi_{\rm d}\rho_{\rm c} r_{\rm c}$, the column density 
along the line of sight $N_{\rm H}$ is given by $N_{\rm H}=\bar{\tau}/(\chi_{\rm d}m_{\rm p})$, where 
$\rho_{\rm c}$ and $r_{\rm c}$ are the density and size of clouds, respectively. The optical depth $\bar{\tau}$ is related with $N_{\rm H}$ by 
$N_{\rm H}\simeq 6\times 10^{21}\bar{\tau}\,{\rm cm}^{-2}$. 
Thus, our model predicts that $f_{\rm obs}$ increases with increasing column density $N_{\rm H}$. Thus, our model indicates 
that $f_{\rm obs}$ becomes larger for higher column density, $N_{\rm H}$, 
e.g., $f_{\rm obs}\sim 0.4$, $0.6$ and $0.8$ for $N_{\rm H}=6\times 10^{21}\, {\rm cm}^{-2}$, $6\times 10^{22}\, {\rm cm}^{-2}$ 
and  $6\times 10^{23}\, {\rm cm}^{-2}$, respectively, which is consistent with X-ray observations (Mateo et al. 2016).
In addition, because the peak AGN luminosity $L_{\rm AGN,p}$ increases as  $N_{\rm H}$ increases, 
the typical BH mass $M_{\rm BH, t}$ decreases with increasing $N_{\rm H}$ (see eq. (19)).  
Note that the dependences of $\bar{\tau}$ is not significant for $M_{\rm BH} > M_{\rm BH, t}$ because the maximal 
$f_{\rm obs}$ is limited  by $C_{\rm *, max}$ (see Fig. 6). 

Lastly, we investigate how $f_{\rm obs}$ depends on the surface density 
of CNDs ($\Sigma_{\rm g}$) for a given $M_{\rm BH}$. Figure 9 shows the case for a gravitationally unstable CND with $\Sigma_{\rm g}=10\Sigma_{\rm g, crit} 
\simeq 30\,{\rm g\, cm}^{-2}$ whose outer radius is $r_{\rm out}=9.5\,{\rm pc}$ obtained by eq. (11).
The figure indicates that, when the surface density of CNDs increases, $f_{\rm obs}$ decreases (i.e., the maximal value of $f_{\rm obs}$ is 0.2),  
and $L_{\rm AGN, p}$ increases because a larger $\Sigma_{\rm g}$ 
results in a lower scale height of CNDs, $h(r_{\rm out})/r_{\rm out}$, owing to the strong gravitational field of CNDs. 
From eqs. (11) and (17), we find  $h(r_{\rm out})/r_{\rm out} \propto \Sigma_{\rm g}^{-9/8}$. 
  
\section{Discussion}
\subsection{Comparison with infrared observations}
Our results on $f_{\rm obs}$ can be compared with the mid-infrared observations of AGNs to check if our predictions reflect the observed structures of the dusty CNDs at $r =1-10$ pc.
Recently, Ichikawa et al. (2019) examined the dust-covering factor of AGNs ($f_{\rm obs, IR}$) 
by using the IR (3--500~$\mu$m) spectral energy distribution for nearby AGNs detected in the all-sky 70-month \textit{Swift}/BAT 
ultra-hard X-ray ($E>10$ keV) survey. 
Their sample contains $\sim600$ AGNs with a wide AGN luminosity range of $10^{41} \,{\rm erg\, s^{-1}}
< L_\mathrm{AGN} < 10^{47}\,{\rm erg\, s^{-1}}$ (the median value is 
$L_\mathrm{AGN} \sim 10^{44.7}\,{\rm erg\, s^{-1}})$ and with a BH mass range of $10^6\,M_{\odot} < M_\mathrm{BH} <
10^{10}M_{\,{\odot}}$ (the median is $M_\mathrm{BH} \sim 10^{8.1}\,M_{\odot}$); 
these values have been obtained from intensive X-ray and optical spectroscopic follow-up observations (Ricci et al. 2017; Koss et al. 2017). 
They found that the dust-covering factor is almost constant with the value $f_\mathrm{obs, IR} \sim 0.5$ 
in the AGN luminosity range of $10^{43}  \,{\rm erg \, s^{-1}}  < L_\mathrm{AGN} < 10^{46} \,{\rm erg\, s^{-1}}$.
Here, we select 179 AGNs with $M_{\rm BH}=10^{7.5}-10^{8.5}M_{\odot}$ from the total of 587 objects in Ichikawa et al. (2019). 

Figure 10 compares the observed data with the hybrid model (red solid line) and SN-driven turbulence model 
(red dashed line) with $M_{\rm BH}=10^{8}M_{\odot}$, $C_{*, {\rm max}}=10^{-7}\,{\rm yr}^{-1}$, $r_{\rm in}=1\, {\rm pc}$, $\bar{\tau}=10$ 
and $\Sigma_{\rm g}=\Sigma_{\rm g, crtit}\simeq 1.0\,{\rm g\, cm}^{-2}$.  
The flat feature around $L_{\rm AGN}\simeq10^{44}-10^{45}\,{\rm erg\,s^{-1}}$ is quantitatively consistent with both models. 
This is also the case for $r_{\rm in}=0.3\,{\rm pc}$ (dot-dashed line in Fig. 10) and $\bar{\tau}=10^2$ (dashed line in Fig. 10). 
In this case, the observed maximum value of $f_{\rm obs}\sim 0.5$  is determined by $C_{*, {\rm max}}$. 
In addition, as mentioned in $\S 2.5$, if $r_{\rm in}$ decreases, the low-luminosity end of  the flat region 
becomes lower. When the optical depth $\bar{\tau}$ increases, the high-luminosity end of the flat feature becomes higher. 
The flat feature does not change significantly, even if we change $r_{\rm in}$ and $\bar{\tau}$, as shown in Fig. 10. 
Moreover, it seems that the hybrid model (SN $+$ radiation pressure model) with $r_{\rm in}=0.3\, {\rm pc}$ well reproduces all data points, 
while the SN-driven turbulence model cannot explain the data of bright AGNs with $L_{\rm AGN}\sim 10^{46}\, {\rm erg}\, {\rm s}^{-1}$. 
To distinguish between two models clearly, it would be important to reduce the error bars of data  
at the lowest and highest luminosity bins (two blue symbols with dashed lines) by increasing the number of objects.

Lastly, we mention the covering factor of obscured quasar. Assef et al. (2015) reported that half of bright quasar  
seem to be obscured by investigating the nature of of hot dust-obscured galaxies selected with the Wide-field Infrared Survey Explorer (\textit{WISE}), 
which are selected hot dust-obscured galaxies. In the present model, 
the obscuring fraction is maximally 0.2 for the parameter range of bright quasars ($M_{\rm BH}\simeq 10^{9}M_{\odot}$) 
and $L_{\rm AGN}/L_{\rm Edd}\simeq 10^{-2}-10^{-1}$ (see Fig. 3). 
This discrepancy implies that the obscuration of bright QSOs may be caused by the gas in their host galaxies ($ >\, 100$ pc) 
and/or highly disturbed $<\,100\,{\rm pc}$ CNDs formed by major mergers.  
For confirmation, it is necessary to observe the dusty-gas distribution of these obscured quasars with ALMA. 

\subsection{Comparison with X-ray observations} 
We compared our results with the luminosity-dependent obscuration in X-ray observations. 
Recent X-ray spectral surveys based on large samples showed that the fraction of obscured AGNs 
peaks ($f_{\rm obs, X} \sim 0.7$) around $L_{\rm X} \sim 10^{43}\,{\rm erg}\,{\rm 3}^{-1}$  (e.g., Burlon et al. 2011; 
Brightman \& Nandra 2011; Buchner et al. 2015; Buchner \& Bauer 2017). 
As shown in Fig. 4 and Fig. 5, the observed $f_{\rm obs}$ for nearby AGNs is consistent with the theoretical predictions 
for $10^7M_{\odot} < M_{\rm BH} < 10^{8}M_{\odot}$. Here, we assume $L_{\rm X}=0.01 - 0.1L_{\rm AGN}$ (e.g., Marconi et al. 2004). 
In addition, Ricci et al. (2017) suggested that the obscuring fraction decreases with the Eddington ratio 
in the range $L_{\rm AGN}/L_{\rm Edd} > 10^{-2}$. This trend appears for a wide range of $M_{\rm BH}$ in Fig. 4 and Fig. 5. 
Thus, these X-ray observations suggest that the obscuring structure is produced by the SN feedbacks 
at low AGN luminosities, i.e., low star-formation efficiencies, while the geometry of the obscuring CND is regulated 
by the AGN radiative feedback; in other words, the gas clouds at high altitude are expelled by the radiation pressure from AGNs 
in the regime of high Eddington ratio. 

However, the covering factor observed in X-rays (Ichikawa et al. 2019) is larger ($f_{\rm obs, X}$) than our predictions 
for any $L_{\rm AGN}$, as shown in Fig. 10. Ricci et al. (2017) also suggested the observed $f_{\rm obs}$ is almost 
constant ($f_{\rm obs, X}\sim 0.7$) between $L_{\rm AGN}/L_{\rm Edd}=10^{-4}$ and $10^{-2}$. 
This discrepancy could be solved if the star formation efficiency ($C_{*}$) assumed in our model is larger, 
because of $h(r_{\rm out})/r_{\rm out}\propto C_{*}^{1/2}$ (see eq. (9)). For example, the maximal $f_{\rm obs}$ 
becomes 0.65, compared to 0.5 in the fiducial case if we assume the high star formation efficiency ($C_{*}\,=10^{-6}\,{\rm yr}^{-1}$) 
as observed in high-$z$ luminous QSO hosts (e.g., Walter et al. 2004; Izumi et al. 2018). The difference between $f_{\rm obs, IR}$ 
and $f_{\rm obs, X}$ may suggest that there are multiple components in CNDs, 
i.e., the layer of the X-ray absorbers (gas+dust) is located above that of 
IR absorbers, because the IR absorbers with higher density is hard to puff up by the SN feedbacks as shown in Figure 11 (a) 
(see also Wada 2015; Wada et al. 2016).  Using ALMA, Izumi et al. (2018) found that the torus in the Circinus galaxy has different scale heights 
in the atomic and molecular gas; The less dense atomic gas forms a thicker disk. This kind of stratified structure may explain 
the dust deficient absorber.
An alternative possibility is that the obscuring structures at optical/IR and X-ray bands are intrinsically different as shown in Figure 11 (b), 
i.e., the covering angle of the dust-free gas structure inside the dust sublimation radius is larger than that of the dusty CND 
 (e.g., Merloni et al. 2014; Davies et al. 2015; Ichikawa et al. 2019).  If this is the case, the dust-free absorbers inside the dusty 
structure ($< 1$pc) would be an essential structure to determine the obscuring fraction ($f_{\rm obs, X}$) for AGNs with 
low Eddington ratios of $L_{\rm AGN}/L_{\rm Edd} < 10^{-2}$. However, the origin of that component is not  clear.

\section{Summary}
We investigated the structure of 10 pc-scale obscuring CNDs
by considering the SN feedbacks from nuclear starburst and the effect of
anisotropic radiation pressure. We explored how structures of 1--10 pc dusty CNDs
depend on the BH mass ($M_{\rm BH}$), AGN luminosity ($L_{\rm AGN}$), and physical properties of CNDs.
Our findings are summarized as follows:

\begin{itemize}

\item
The obscuring fraction, $f_{\rm obs}$,  peaks at the luminosity $L_{\rm
AGN, p}\sim 10\%$ of  the AGN Eddington luminosity ($L_{\rm Edd}$),
and  the maximal value of  $f_{\rm obs}$ is $\sim0.6$ for less massive
SMBHs (e.g., $M_{\rm BH} < 10^{8}M_{\odot}$).
For lower $L_{\rm AGN}$, the obscuring fraction is determined by the
SN feedback, while the radiative feedback is important for higher $L_{\rm AGN}$.
On the other hand, for massive SMBHs (e.g., $M_{\rm BH} >10^{8}M_{\odot}$), 
the obscuring fraction $f_{\rm obs}$ is always low ($< 0.2$), and 
it is independent of $L_{\rm AGN}$ because the scale height of CNDs
is mainly regulated by the maximal star-formation efficiency, $C_{*. \rm{max}}$,
in CNDs.

\item The maximal $f_{\rm obs}$ slightly increases as the inner radius
of CNDs ($r_\mathrm{in}$) decreases. This case may correspond to 
heavily obscured AGNs with relatively low-mass BHs ($M_{\rm BH} < 10^{8}M_{\odot}$).
In addition, our model indicates that $f_{\rm obs}$ increases with the column density 
of line of sight $N_{\rm H}$, which is consistent with recent X-ray observations 
(Mateo et al. 2016). Moreover, when the surface density of CNDs is larger, $f_{\rm obs}$
is smaller (i.e., the maximal value of $f_{\rm obs}$ being 0.2), 
and $L_{\rm AGN, p}$ becomes larger. We then predict that $f_{\rm
obs}$ decreases with the surface density of the obscuring 
materials.

\item We compared the predicted obscuring fraction $f_{\rm obs}$ with
mid-IR observations (Ichikawa et al. 2019). 
The SN $+$ radiation pressure model is consistent with the IR
obscuring fraction for massive BHs with $M_{\rm BH}=10^{8}M_{\odot}$.  
This implies that an intense nuclear starburst with $C_{*, {\rm max}}=
10^{-7}\,{\rm yr}^{-1}$ contributes to the obscuration in these objects. 
In addition, our model can qualitatively explain the observed behavior
of $f_{\rm obs}$ as a function of the X-ray luminosity (e.g., Burlon
et al. 2011).  
However, $f_{\rm obs, X}$ is always greater than our theoretical predictions, especially for AGNs 
with low Eddington luminosity ratio ($L_{\rm AGN}/L_{\rm Edd} < 10^{-2}$).
One solution is the high star formation efficiency ($C_{*}\,=10^{-6}\,{\rm yr}^{-1}$) 
as observed in high-z luminous QSO hosts. The other option is the  major contribution of 
the dust-free absorbers inside  the dust sublimation radius in the CNDs ($< 1$pc).  

\end{itemize}

As mentioned above, the current model cannot explain the dust-free obscuring structure 
for AGNs with low Eddington ratio,  $L_{\rm AGN}/L_{\rm Edd} < 10^{-2}$. 
To resolve this issue, it might be important to take into account a failed dusty wind from the outer accretion disk 
(e.g., Czerny \& Hryniewicz 2011; Baskin \& Laor 2018) because this effect works at the dust-free region inside dusty CNDs.  
Furthermore,  in this work, we considered how both the SN and radiative feedbacks from AGNs affect on 
the obscuring structure of AGNs. The mechanical feedbacks by strong AGN outflows (e.g., Nomura \& Ohsuga 2017) 
may also be important for the obscuring fraction of AGNs, because high-velocity outflows 
with the velocity of $10\%$ of speed of light are detected in  almost half of 
Seyfert galaxies (e.g., Tombesi et al. 2010, 2011; Gofford et al. 2013). 
The effect of AGN winds on $f_{\rm obs}$ will be left in our future work.

\acknowledgments 
We thank an anonymous referee for scientific suggestions that helped improve the paper.
We are very grateful to Imanishi M., Toba. Y.  and Izumi, T. for useful comments and discussions.
NK acknowledges JSPS   KAKENHI   Grant   Numbers 16K17670 and  19K03918. 
This  work  was  supported  by JSPS  KAKENHI  Grant  Number 16H03958 (KW). 
This study also benefited from financial support from JSPS  KAKENHI  Grant  
Number 18K13584 (KI) and the Japan Science and Technology Agency (JST) grant
 ``Building of Consortia for the Development of Human Resources in Science and Technology'' (KI).

\begin{figure}
\plotone{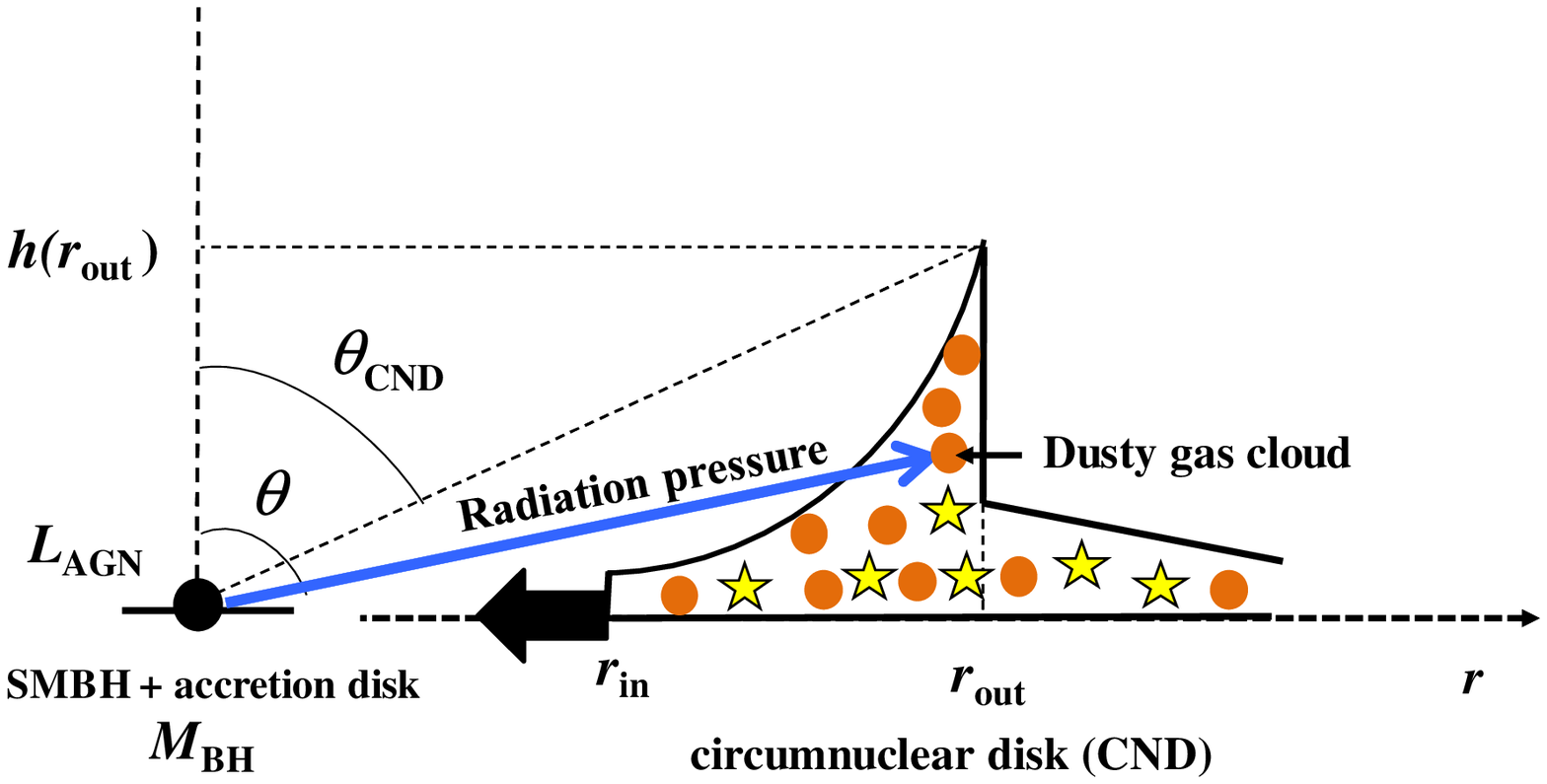}
\caption
{Schematic view of a circumnuclear disk (CND) and the effect of radiation pressure from the AGN.  
The angle between the line of sight and the normal to the accretion disk is defined as  $\theta$. 
The thickness of CND is expressed by $\tan(\frac{\pi}{2}-\theta_{\rm CND})\equiv h(r_{\rm out})/r_{\rm out}$.  
}
\end{figure} 

\begin{figure}
\plotone{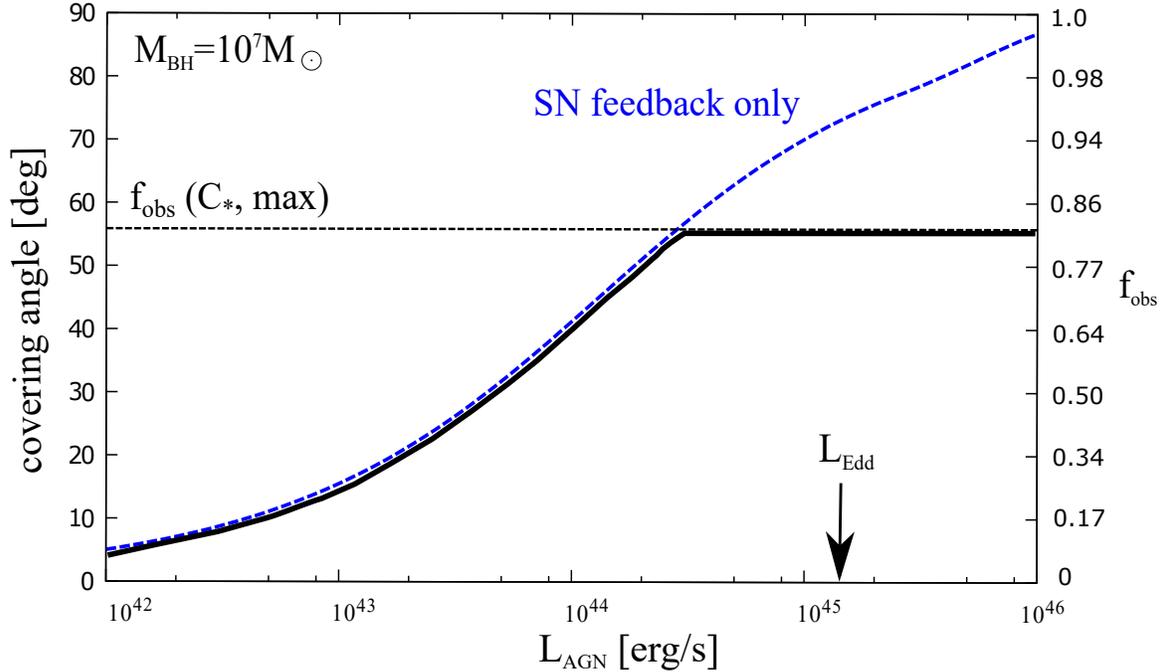}
\caption
{
AGN obscured fraction $f_{\rm obs}$ (right-hand axis of ordinate) and the covering angle $\frac{\pi}{2}-\theta_{\rm CND}$ 
(left-hand axis of ordinate) against the AGN luminosity $L_{\rm AGN}$ for $M_{\rm BH}=10^{7}M_{\odot}$. 
The blue dashed line shows $f_{\rm obs}$ obtained from the SN feedback only (eq. (17)), while the horizontal dashed line 
shows the obscuring fraction $f_{\rm obs}(C_{*, {\rm max}})$ for the maximal star-formation efficiency 
$C_{\rm *, max}=10^{-7}\,{\rm yr}^{-1}$ (eq. (9)). 
The thick black line represents the obscuring fraction ($f_{\rm obs}$) predicted by the SN-driven turbulence model. 
The maximal $f_{\rm obs}$ is $\sim 0.8$ at $L_{\rm AGN} \ge 0.1L_{\rm Edd}$. 
The arrow shows the AGN Eddington luminosity $M_{\rm BH}=10^{7}M_{\odot}$. 
}
\end{figure}

\begin{figure}
\plotone{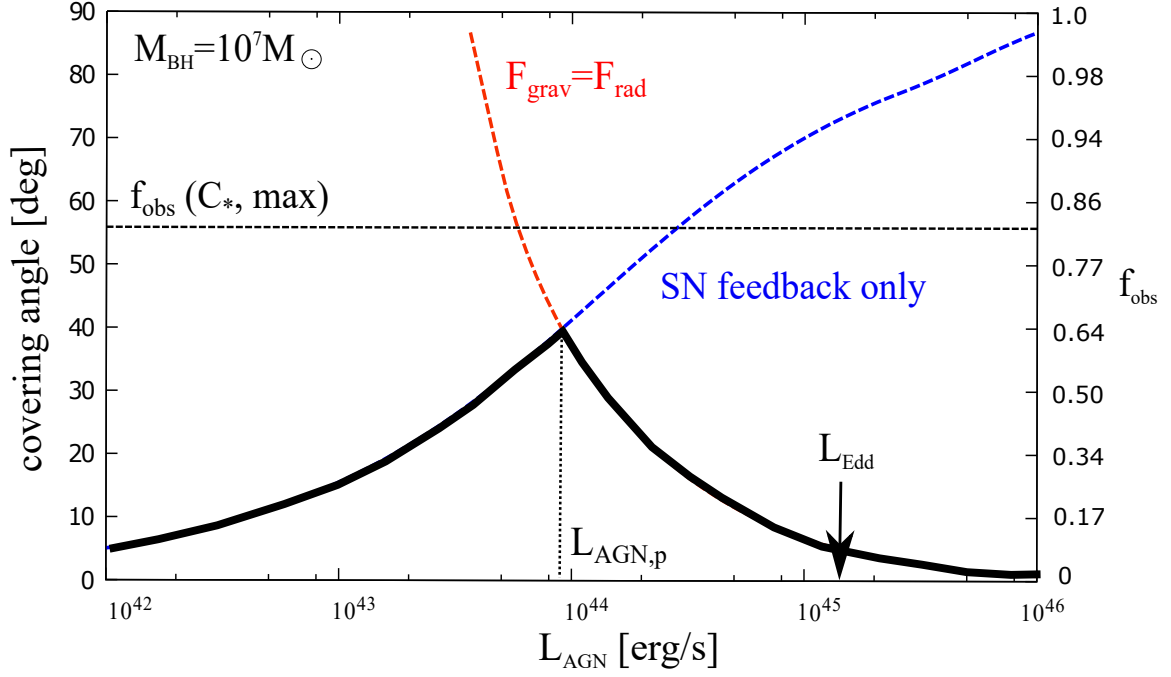}
\caption
{
Same as Fig. 2, but with the effect of anisotropic radiative pressure from AGNs. 
The red dashed line shows  the required luminosity that can balance the gravitational force of SMBHs with $\bar{\tau}=10$ (eq. (12)). 
The thick black line represents the obscuring fraction, $f_{\rm obs}$, predicted by the hybrid model, which considers 
the SN feedback and the radiative feedback from the AGN. 
The maximal $f_{\rm obs}$ is approximately 0.6 at $L_{\rm AGN, p}$. 
}
\end{figure} 

\begin{figure}
\plotone{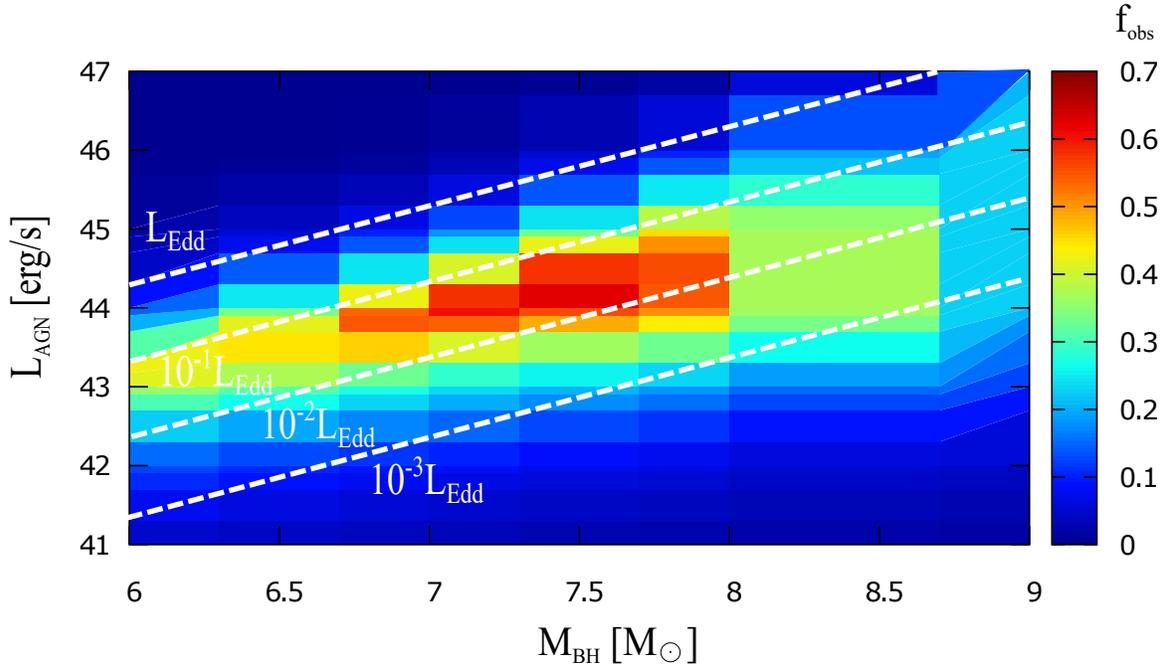}
\caption
{
Contours of the obscuring fraction, $f_{\rm obs}$, of hybrid models (SN+radiative feedback models) 
for various $M_{\rm BH}$ and $L_{\rm AGN}$, assuming $r_{\rm in}=1\,{\rm pc}$, 
$\Sigma_{\rm g}=\Sigma_{\rm g, crit}$, and $\bar{\tau}=10$. 
The four dashed lines represent $L_{\rm AGN}=L_{\rm Edd}$, $10^{-1}L_{\rm Edd}$, $10^{-2}L_{\rm Edd}$, and $10^{-3}L_{\rm Edd}$, 
respectively.
}
\end{figure}

\begin{figure}
\plotone{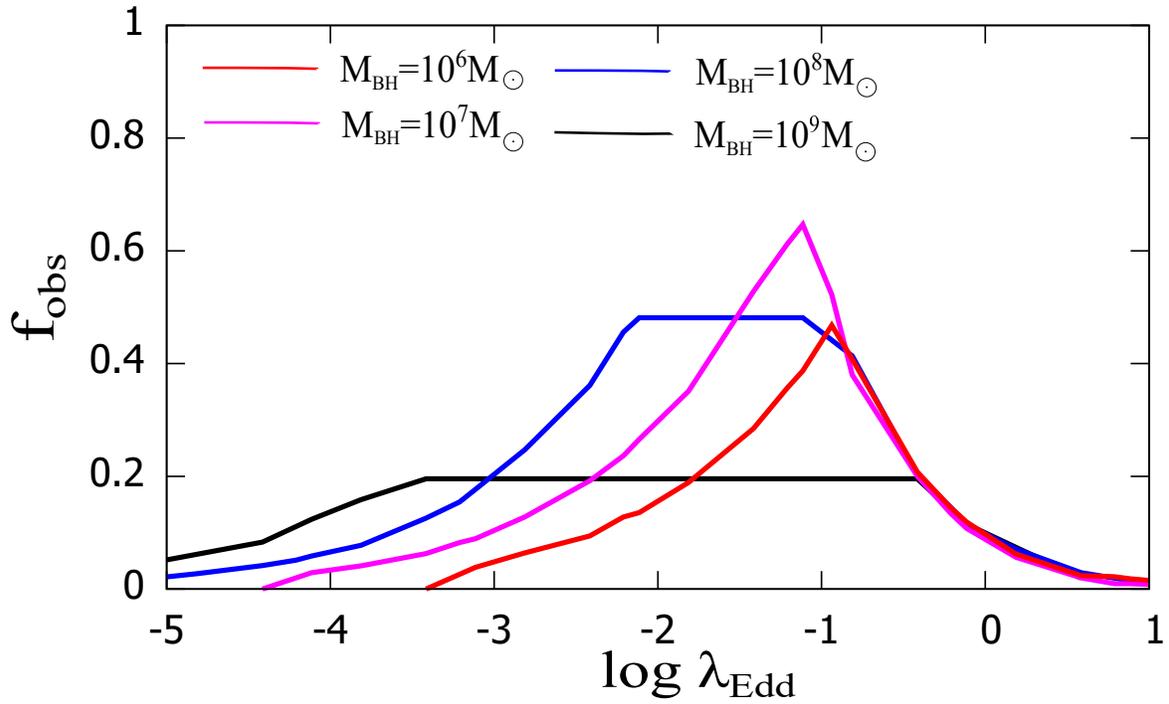}
\caption
{
The AGN obscuring fraction $f_{\rm obs}$ against the Eddington ratios $\lambda_{\rm Edd}=L_{\rm AGN}/L_{\rm Edd}$
for $M_{\rm BH}=10^{6}M_{\odot}$, $10^{7}M_{\odot}$, $10^{8}M_{\odot}$ and $10^{9}M_{\odot}$. 
}
\end{figure}

\begin{figure}
\plotone{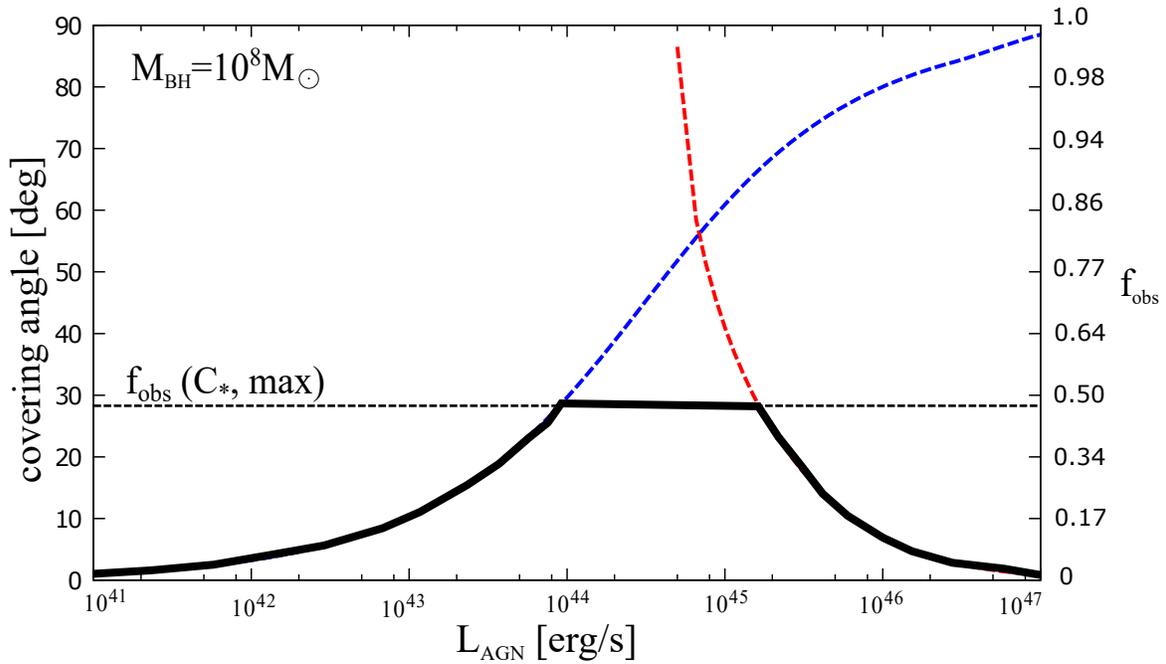}
\caption
{
Same as Fig. 3, but for $M_{\rm BH}=10^{8}M_{\odot}$ 
with $r_{\rm in}=1\,{\rm pc}$ and $r_{\rm out}=53\,{\rm pc}$.
}
\end{figure} 

\begin{figure}
\plotone{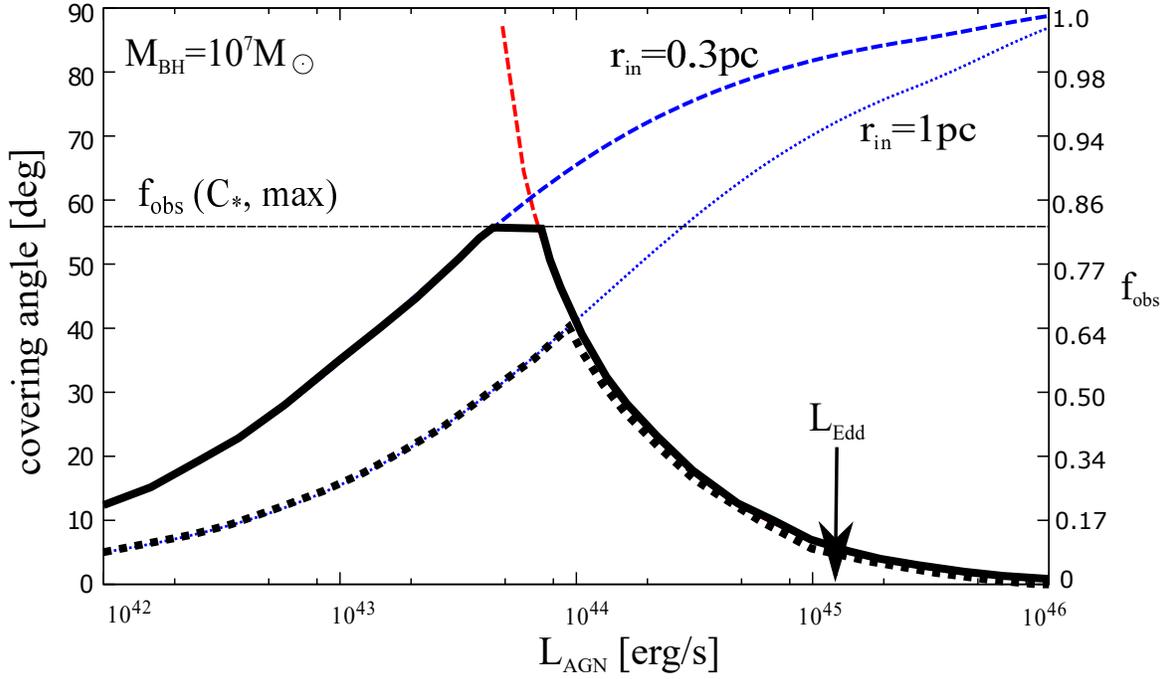}
\caption
{
Same as Fig. 3, but for a smaller inner radius $r_{\rm in}=0.3\,{\rm pc}$ (thick black line). 
The dotted black line corresponds to the case of  $r_{\rm in}=1\,{\rm pc}$. 
}
\end{figure} 

\begin{figure}
\plotone{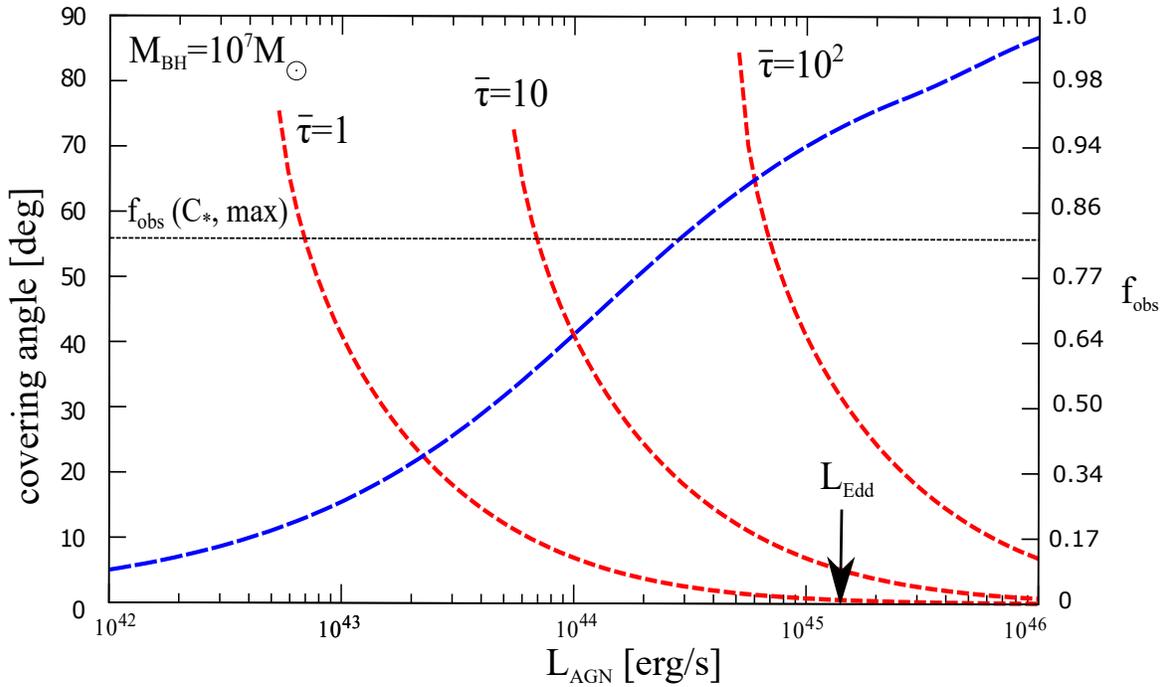}
\caption
{
Same as Fig. 3, but for different optical depths of gas clouds 
$\bar{\tau}=1, 10$, and $10^2$.
}
\end{figure} 

\begin{figure}
\plotone{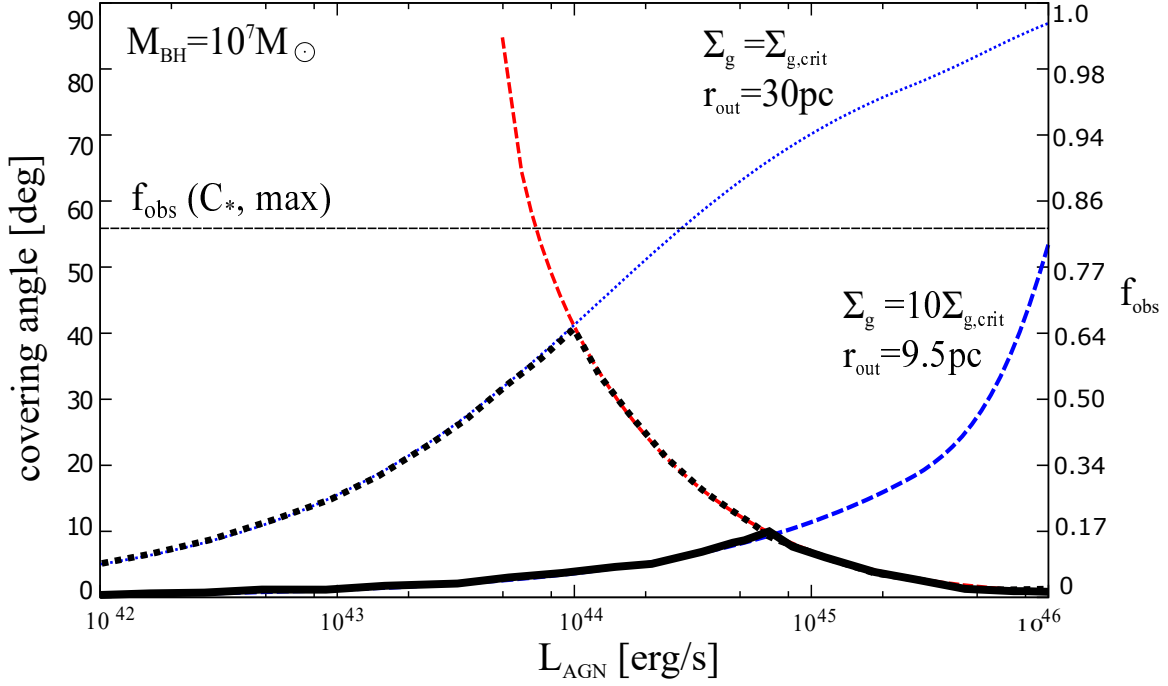}
\caption
{
Same as Fig. 3, but for a higher surface density $\Sigma_{\rm g}=10\Sigma_{\rm g, crit}\simeq 30\,{\rm g\, cm}^{-2}$ (thick black line). 
The dotted black line corresponds to the case of  $\Sigma_{\rm g}=\Sigma_{\rm g, crit}\simeq 3.0\,{\rm g\, cm}^{-2}$. 
}
\end{figure} 

\begin{figure}
\plotone{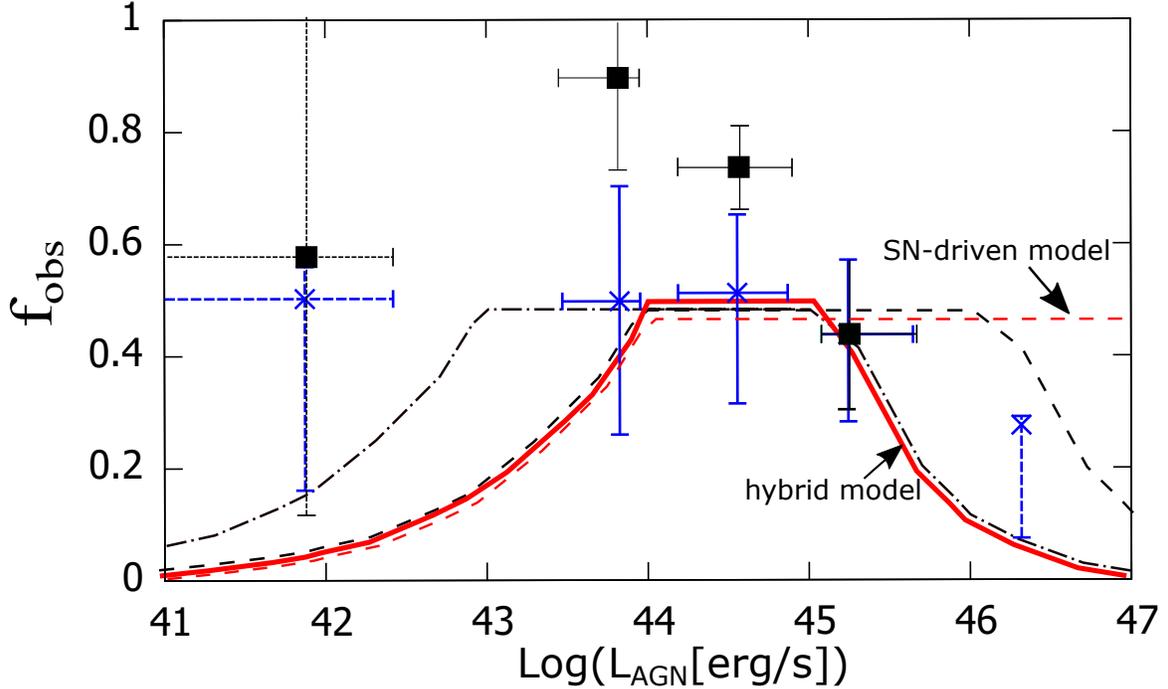}
\caption
{
Comparison with IR observational data (blue symbols)  and X-ray data (black symbols) of nearby AGNs 
(both IR and X-ray data from Ichikawa et al. 2019) with the average BH mass $\simeq 10^8M_{\odot}$. 
The red solid line represents the prediction of the hybrid model (SN $+$ radiation pressure model) 
with $M_{\rm BH}=10^{8}M_{\odot}$,  $C_{\rm *, max}=10^{-7}\,{\rm yr}^{-1}$, 
$r_{\rm in}=1\, {\rm pc}$, $\bar{\tau}=10$ and $\Sigma_{\rm g}=\Sigma_{\rm g, crtit}\simeq 1.0\,{\rm g\, cm}^{-2}$. 
The SN-driven turbulence model is shown by the red dashed line (see also Fig. 2). 
The black dashed line corresponds to a larger optical depth $\bar{\tau}=10^2$, while the black dotted-dashed line 
corresponds to a smaller inner radius $r_{\rm in}=0.3$ pc. 
}
\end{figure} 

\begin{figure}
\plotone{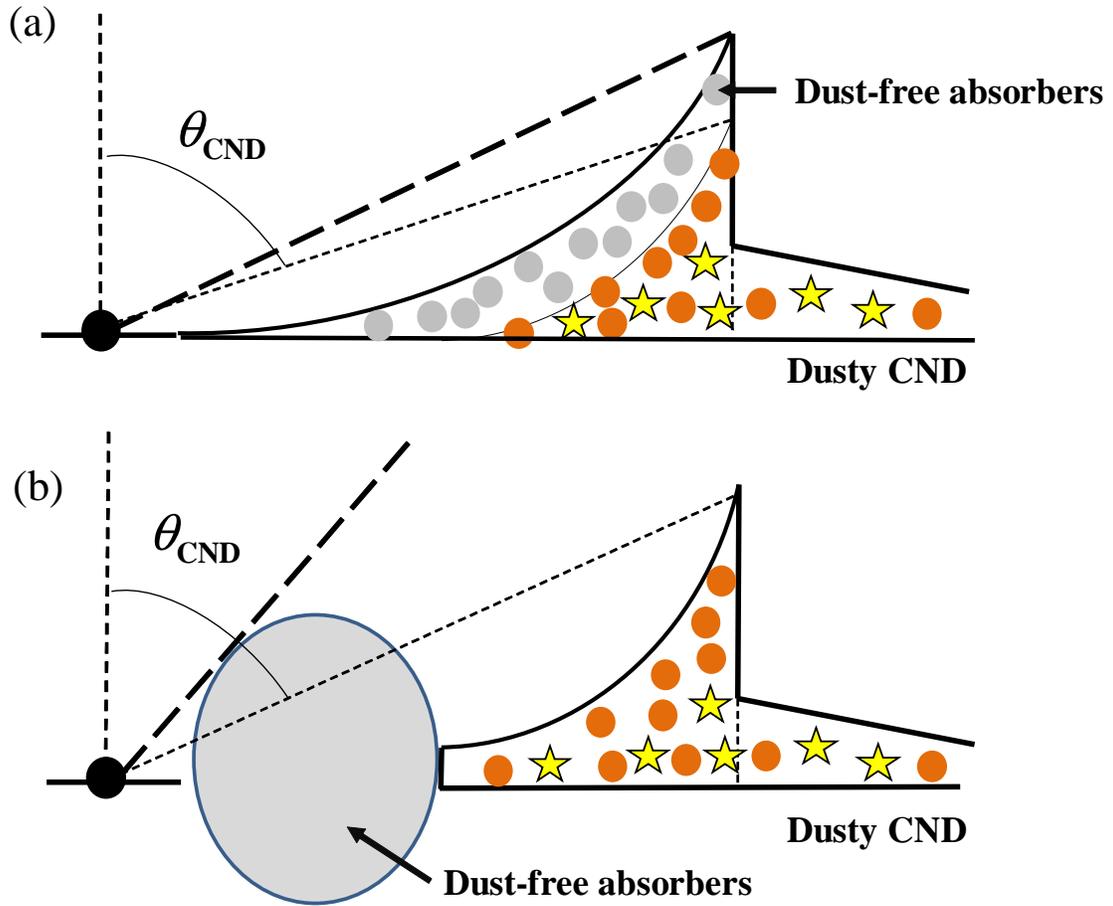}
\caption
{
Schematic pictures for two scenarios to explain the discrepancy between the X-ray observations and our theoretical model. 
The case (a)  corresponds to the intense nuclear starburst with $C_{\rm *}=10^{-6}\,{\rm yr}^{-1}$ and multiphase CNDs. 
The case (b) describes that the dust-free absorbers inside the dust sublimation radius in the CNDs ($< 1$pc) contributes 
the X-ray observations, $f_{\rm obs, X}$.
}
\end{figure}

\end{document}